# Thermal Management of Lithium-Ion Batteries: A Comparative Study of Phase Change Materials and Air-Cooling Systems Equipped with Fins

Masoumeh Karimi Kisomi

Department of Mechatronics Engineering, Technische Universität Ilmenau, Ilmenau, 98693, Germany.

**Abstract**

Due to the limitations of conventional energy sources and concerns regarding environmental pollution, battery-powered vehicles have garnered widespread attention. Lithium-ion (Li-ion) batteries are extensively utilized as the primary power source for electric vehicles due to their high energy density, environmental friendliness, long lifespan, and lightweight nature. However, their performance and safety are highly dependent on operating temperature. Therefore, a battery thermal management system (BTMS) is essential to ensure the reliable operation and safety of electric vehicles. This study presents a battery thermal management system incorporating phase change material (PCM) and air cooling in a cylindrical lithium-ion cell with fins to enhance heat dissipation. This research examines PCM and air cooling systems to enhance battery heat transfer performance. The effects of each system on maximum temperature, minimum temperature, and temperature uniformity along the battery cell are analyzed. Additionally, the impact of fins in both systems is evaluated against a finless cell. A numerical analysis utilizing ANSYS software and the finite volume method (FVM) is performed to evaluate the cooling performance of the systems. The results show that PCM reduces both the maximum and minimum temperatures compared to the air cooling system due to the phase change mechanism. In the finless battery case, the maximum temperature decreases from 316 K to 304 K when using PCM instead of the air cooling system. Also, in the same fin-based battery, the minimum temperature decreases from 307 K to 302 K by using PCM instead of the air cooling system, leading to improved temperature stability. The results indicate that, in general, the fins help reduce the maximum cell temperature when compared to the case without fins in both cases. Using rectangular fins reduces the maximum temperature by approximately 3% compared to a finless battery in the air cooling system. Additionally, the presence of fins reduces the temperature difference along the battery, ensuring a more uniform temperature distribution, such that, in the PCM system with rectangular fins, the temperature difference remains below 1 K. Also, the effect of the number of fins was investigated in this study. Increasing the number of fins enhances the heat transfer area, and when the number of square fins increases from 2 to 4, the maximum temperature decreases from 311 K to 309 K.

# 1 Introduction

To reduce energy consumption and carbon emissions, the shift from conventional fuel-powered vehicles to new energy vehicles has become a growing trend. In response to the global energy crisis and the increasing focus on environmental protection and energy efficiency, electric vehicles are emerging as a promising solution to reduce carbon emissions.

Among various battery types, lithium-ion batteries are the preferred power source for electric vehicles due to their long lifespan, low environmental impact, lightweight, high power output, and excellent discharge efficiency [1]. However, overheating and uneven heat distribution within the cells can lead to thermal runaway and battery degradation. Therefore, it is essential to implement an effective BTMS to ensure the safety and reliability of the battery pack. As a result, the design and optimization of BTMS are critical for maximizing the performance and safety of lithium-ion batteries in electric vehicles. The optimal operating temperature range for lithium-ion batteries typically falls between 20°C and 50°C, as performance significantly declines when the temperature exceeds this range [1]. To achieve optimal performance, it is crucial to maintain uniform temperature distribution across the battery pack. The temperature difference between cells within a battery module should be kept within 5°C to prevent overheating and performance degradation [1]. Various methods are used to manage battery temperature, including air cooling, liquid cooling, phase change materials, heat pipes, thermoelectric modules, and combinations of these approaches [2]. Liquid cooling systems are highly efficient due to the superior heat transfer properties of liquids, but they are often complex in design, prone to leakage, and come with high maintenance costs [3]. While heat pipe cooling methods are also effective, their practical use is limited by high costs and relatively immature technology [3]. PCM is a key component of passive BTMS. During its phase change process, it absorbs heat and helps maintain temperature stability, which aids in extending the temperature regulation period of the BTMS and promoting uniform battery temperature. On the other hand, air-cooled method has attracted significant interest because of their lightweight design, cost-effectiveness, and ease of operation and maintenance [2]. However, the low thermal conductivity of air means that the thermal efficiency of air-cooled systems still needs improvement [2]. Air conditioning systems are particularly favored for their cost-effective nature, simple structure, and leak-free operation. Based on research, each model has its own advantages and disadvantages, depending on the application, working temperature, load conditions, and other factors. In recent years, substantial research has been dedicated to studying the performance of air-cooled BTMS in managing battery temperatures. Several studies have been conducted on the impact of pack's geometry design on battery performance. Suryavanshi and Ghanegaonkar [4] conducted a numerical study on structural design using air cooling method, aiming to introduce an innovative flow circulation technique for uniform airflow distribution. They developed 3D models of nine aluminum perforated plates with different topologies. The study examined the effects of air velocity, inlet temperature, C-rate (charge / discharge current divided by the nominally rated battery capacity), and cell spacing on the nine-plate structure. Their findings indicated that modifying the flow distribution layout significantly enhances the battery pack's cooling performance. In their case, optimal cooling was achieved with a 2 mm cell spacing, ensuring uniform airflow and improved heat dissipation. The best thermal performance was observed at air speeds of 0.8 m/s for 0.5C, 5 m/s for 1C, and 30 m/s for 2C. Chen et al. [5] conducted both experimental and numerical investigations on a parallel air-cooled system. They developed a control strategy for the efficient cooling of battery packs under varying operating conditions. The control strategy is based on the temperature difference between the battery cells. Multiple outlet channels were designed, and valves at the outlets are regulated using this strategy to switch the airflow direction. The control strategy successfully reduces the temperature difference between battery cells by dynamically adjusting the airflow, directing more cooling air toward the hotter cells. The average temperature difference among the battery cells in the developed system is reduced by more than 67% compared to a system using only J-type flow. Yan et al. [6] conducted a numerical study on the effect of structural parameters in a battery pack using the air-cooled method. Their focus was on the design of the number, size, and location of the inlet and outlet openings. The results revealed that when there is only one inlet and outlet, increasing their size improves the heat dissipation performance. When the diameters of the inlet and outlet increase from 120 mm to 210 mm, the maximum temperature difference in the battery module decreases from 20.4°C to 10.7°C. This improvement is primarily due to the higher convective heat transfer coefficient resulting from the larger volume of incoming air. Furthermore, when the air volume remains constant, the optimal heat dissipation performance is achieved with two inlets and outlets. Zhang et al. [7] conducted a numerical study on liquid cooling, focusing on a microchannel plate positioned at the bottom with added columns featuring three arc-shaped sides attached to cylindrical batteries. They analyzed the effects of column height and coolant flow rate on the system's maximum temperature and temperature difference. Their findings indicated that the coupled cooling model reduced the maximum temperature by 16.6% and 32.0% at 1C and 2C discharge rates, respectively, compared to conventional cooling. Additionally, increasing the thermal column height led to a decrease in both maximum temperature and temperature difference. However, when the column height exceeded 60 mm, further increases became unnecessary, as they compromised the system's mass energy density without significant thermal improvement.

Several studies have explored the geometry of battery walls and analyzed their impact on performance. Gao et al. [8] performed a numerical study on the geometric design of an air-cooled BTMS, focusing on mitigating the high temperatures of downstream batteries, which are influenced by the airflow from upstream. They introduced flow splitters positioned at the back of the batteries to enhance the thermal-hydraulic performance. The inclined flow splitters helped reduce the maximum temperature and improve temperature uniformity by promoting heat conduction and minimizing the wake zone size. With the inclined flow splitter, the maximum temperature of the battery decreased by 2.14°C, and the temperature difference was reduced by up to 49.2%, all while maintaining the same pumping power as the original model without the flow splitter. Also, researchers have worked on phase change materials for battery thermal management systems as a passive solution that does not require an external power source for cooling. Maher Abd et al. [9] conducted experimental research on a hybrid BTMS that combines phase change materials with aluminum fins and forced air to improve battery cooling performance. They investigated the effects of various discharge rates and air velocities. The results showed that the hybrid model significantly reduces heat buildup in the battery, lowering the maximum operating temperature by 1.5°C, 5.5°C, and 9.5°C compared to the air-cooled model, and by 2.8°C, 5.1°C, and 16.1°C compared to the PCM-only model for discharge rates of 1C, 2C, and 3C, respectively. Luo et al. [10] investigated the thermal performance of two configurations: Batteries-PCM and batteries-PCM-fins. Their findings reveal that incorporating snowflake-shaped fins in the Batteries-PCM-Fins design effectively reduces battery temperatures at a 3C discharge rate, keeping the maximum temperature difference below 3°C. Ultimately, they proposed an optimized design. In this configuration, when the coolant flow rate is 0.1 m/s and the discharge rate is 5C, the maximum battery module temperature is reduced by 17.40% at 25°C ambient temperature and by 21.36% at 40°C compared to a design without a liquid cooling plate. Additionally, the maximum temperature difference decreases by 42.53% and 55.77%, respectively. Heyhat et al. [11] studied the performance of lithium-ion batteries that incorporated nanoparticles, fins, and porous metal foam in addition to PCM. Their results showed that the porous-PCM combination performed more efficiently than both nano-PCM and fin-PCM configurations. Specifically, the use of porous-PCM resulted in a 4–6 K reduction in the battery's mean temperature compared to pure PCM. However, for the porous-PCM composition, a delay in the initiation of PCM melting was observed, which could negatively impact the performance of the battery's thermal management system.

In this study, a comparative analysis is conducted between an air-cooled system and a PCM-based system under identical operating conditions. Additionally, various fin designs are integrated into the battery to enhance heat transfer and determine the most suitable configuration. To achieve this, a single cylindrical battery is first simulated using an air-cooled thermal management system, beginning with a finless configuration. The study then examines and compares four fin geometries circular, square, and rectangular to determine the most effective design for preventing thermal runaway and ensuring battery safety under extreme conditions. The most efficient fin configuration is then integrated into the PCM-based system to evaluate its effectiveness compared to the air-cooled system. The thermal performance of each system is evaluated based on the maximum temperature and temperature variation within the battery. The finite volume method is employed for numerical analysis. The key objectives of this study are: (a) to compare the performance of air-cooled and PCM-based systems under identical conditions, (b) to investigate the influence of fins on the heat transfer characteristics of the battery thermal management system, and (c) to identify the optimal fin design for achieving efficient thermal management of batteries.

## 2 Numerical Study

### 2-1 Geometry

This study is conducted using a 3D simulation model. The battery pack consists of one cylindrical cell, as shown in Fig. 1. The enclosure and cell dimensions are provided in Table 1. The fins, made of aluminium due to its low cost and lightweight properties, are attached to the battery surface and extend into the surrounding air to enhance heat dissipation. The geometry of the fins is shown in Fig. 2. In this study, circular, square, and rectangular shapes are considered as possible fin designs.

(a)            (b)

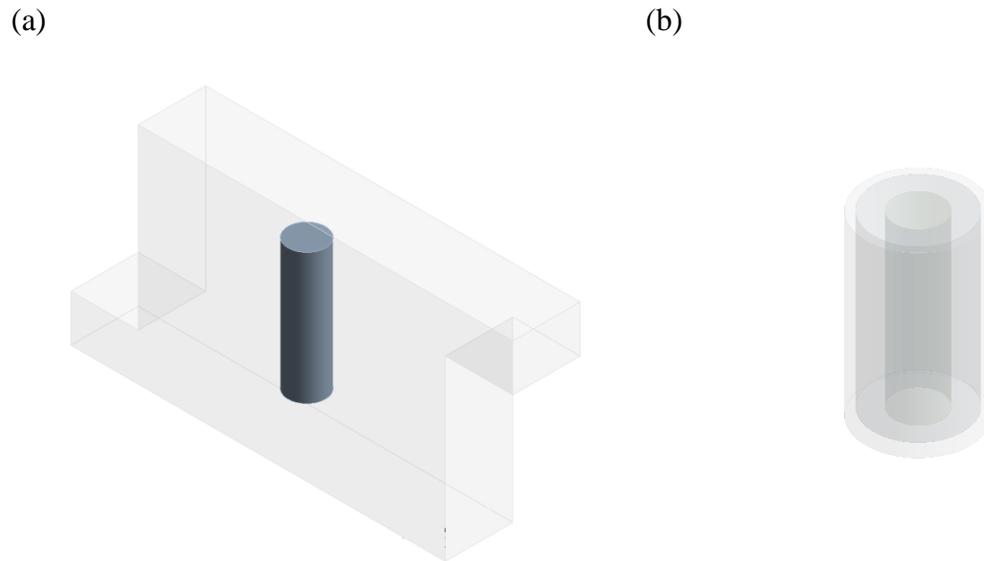

**Fig. 1** (a) Configuration of the single battery cell and enclosure in the air-cooled system and (b) Battery cell in PCM system.

Table 1. Geometric parameters of battery and enclosure.

| Geometry parameter | Value (mm) |
|---|---|
| Battery diameter | 20 |
| Battery height | 70 |
| Length of channel | 200 |
| Height of channel | 120 |
| Inlet and outlet (square) | 25 |

(a)

(b)

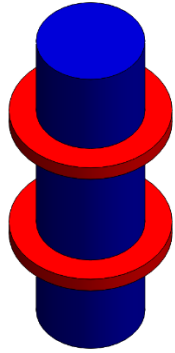
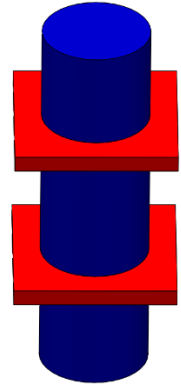

(c)

(d)

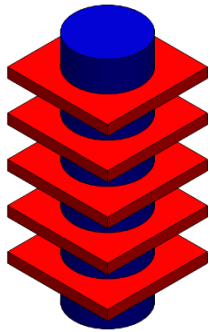
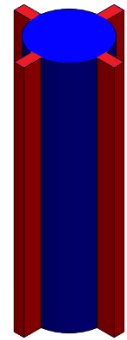

**Fig. 2** Fin geometries: (a) Two circular fins, (b) Two square fins, (c) Four square fins, and (d) Four rectangular fins.

Fig. 3 shows a 2D diagram of the battery with fins, including the side and top views. The corresponding dimensions are provided in Table 2. The thickness of the fins is considered 3 mm. The material properties of the battery, fins, enclosure, and air are listed in Table 3.

**Table 2.** Dimension of fins.

| Parameter | Size (mm) | Parameter | Size (mm) | Parameter | Size (mm) |
|---|---|---|---|---|---|
| $L_1$ | 30 | $L_4$ | 3 | $t_1$ | 3 |
| $L_2$ | 30 | $R_1$ | 10 | $t_2$ | 3 |
| $L_3$ | 7 | $R_2$ | 15 | $H_1$ | 70 |

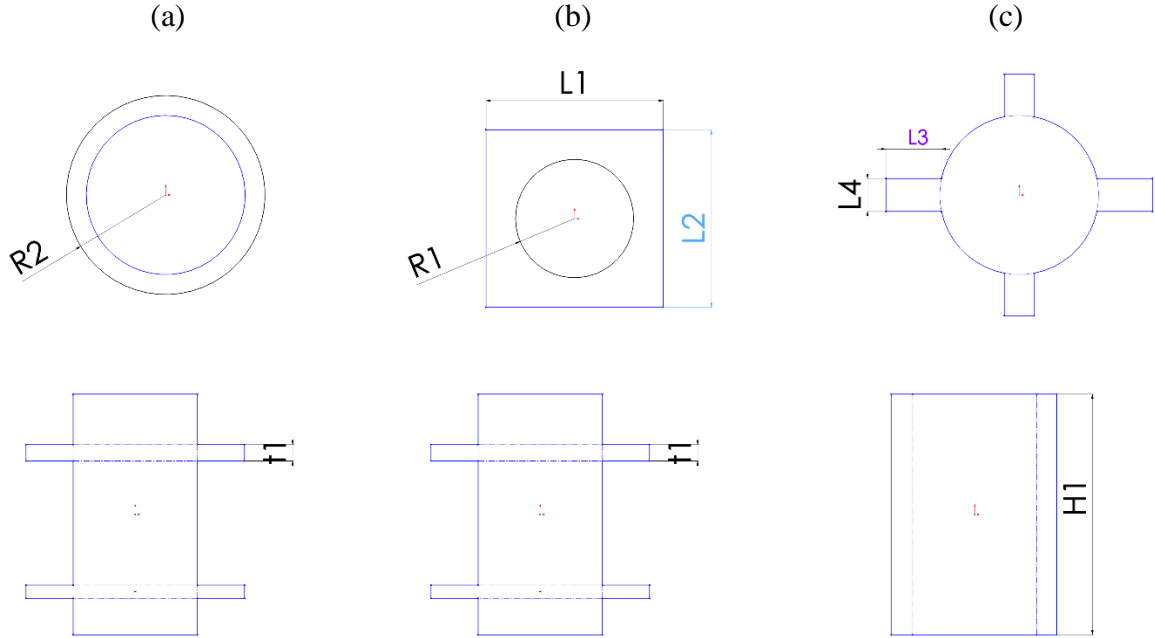

**Fig. 3** 2D schematic of fin designs in air-cooled system: (a) Two circular fins, (b) Two square fins, (c) Four rectangular fins.

**2-2 Air Flow Simulation**

In this study, the finite volume method (FVM) using ANSYS Fluent Software 2023 R2 is utilized. The airflow is modeled by solving the Navier-Stokes equations, with the continuity and momentum equations represented in Eq. (1) and Eq. (2) [12,7]:

$$\frac{\partial \rho}{\partial t} + \frac{\partial (\rho \bar{u}_i)}{\partial x_i} = 0 \tag{1}$$

$$\frac{\partial (\rho \bar{u}_i)}{\partial t} + \frac{\partial (\rho \overline{u_i u_j})}{\partial x_j} = \rho g_i - \frac{\partial P}{\partial x_i} + \frac{\partial \tau_{ij}}{\partial x_j} \tag{2}$$

Where $\rho$ $(kg.m^{-3})$ is air density, $u$ $(m.s^{-1})$ is the velocity of the continuous phase, $P$ $(Pa)$ is pressure of the continuous phase, and $\tau_{ij}(N.m^{-2})$ is Reynolds stress tensor. Based on the Reynolds number, the airflow regime is classified as turbulent, and the $k - \varepsilon$ turbulence model is applied, as shown in Eq. (3) and Eq. (4) [12].

$$\frac{\partial}{\partial t}(\rho k) + \frac{\partial}{\partial \text{xi}}(\rho u_i k) = \frac{\partial}{\partial \text{xi}}[(\mu + \frac{\mu_t}{\sigma_k})\frac{\partial k}{\partial \text{xi}}] + G_k - \rho\varepsilon \tag{3}$$

$$\frac{\partial}{\partial t}(\rho \varepsilon) + \frac{\partial}{\partial \text{xi}}(\rho u_i \varepsilon) = \frac{\partial}{\partial \text{xi}}[(\mu + \frac{\mu_t}{\sigma_k})\frac{\partial \varepsilon}{\partial \text{xi}}] + C_{1\varepsilon}G_k\frac{\varepsilon}{k} - C_{2\varepsilon}\rho\frac{\varepsilon^2}{k} \tag{4}$$

Where $k$ ($m^2.s^2$) is turbulence kinetic energy, $\varepsilon$ ($m^2.s^3$) is dissipation rate of turbulent kinetic energy. $G$ ($m^2.s^3$), $C_{1\varepsilon}$ (-), $C_{2\varepsilon}$ (-), $\sigma_k$ (-) are constants, while $\mu_t$ ($Pa.s$) represents the turbulent viscosity.

## 2-3 Heat Transfer Simulation

Heat transfer in the air is analyzed using the energy equation represented in Eq. (5) [7]:

$$\rho C_p \left(\frac{\partial T}{\partial t}\right) + \rho C_p (u.\nabla T) = \nabla.\left[\nabla(k_{eff}T)\right] \tag{5}$$

Where $T(K)$ represents the air temperature, $C_p$ ($j.kg^{-1}.K^{-1}$) denotes the specific heat capacity of air, and $k_{eff}$ ($W.m^{-1}.K^{-1}$) refers to the effective thermal conductivity. The battery's temperature distribution is determined using Eq. (6) [12]:

$$\rho_b C_{p,b} \left(\frac{\partial T_b}{\partial t}\right) = \frac{\partial}{\partial x_j}\left[k_{b,j}\frac{\partial T}{\partial x_j}\right] + \varphi \tag{6}$$

Where $T_b$ ($k$) represents the battery temperature, $k_b$ ($W.m^{-1}.K^{-1}$) is the battery's thermal conductivity, $C_{p,b}$ ($j.kg^{-1}.K^{-1}$) denotes the battery's heat capacity, and $\varphi$ ($W.m^{-3}$) is the battery's heat generation rate.

In Eq. (6), $\varphi$ represents the thermal source term, corresponding to the heat generation rate of the battery. According to Bernardi et al., the total heat generation in the battery consists of two components: Joule heating and entropy heating. Joule heating occurs when an electric current flows through a conductor, generating heat due to electrical resistance [7]. This process is irreversible and contributes to the majority of the heat produced within the cell. Entropy heating arises from entropy changes during electrochemical reactions inside the battery. Unlike Joule heating, it is a reversible heat generation process associated with the electrochemical activity of the cell. $\varphi$ is expressed as in Eq. (7) [7]:

$$\varphi = I(E_{OCV} - E_{cell}) + IT\left(\frac{\partial E_{OCV}}{\partial T}\right) \tag{7}$$

Where $I$ ($A$) is the discharge current, $E_{OCV}$ ($V$) is the open-circuit voltage, $E_{cell}$ ($V$) is the cell potential, $T$ ($K$) is the ambient temperature, and $\frac{\partial E_{OCV}}{\partial T}$ ($V.K^{-1}$) is the entropy change within a cell.

**Table 3.** Thermal Properties of air, battery, and fins [13].

| Material/property | $\rho$ ($kg.m^{-3}$) | $C_p$($j.kg^{-1}.K^{-1}$) | $k$ ($W.m^{-1}.K^{-1}$) |
|---|---|---|---|
| Air | 1.225 | 1006.43 | 0.0242 |
| Battery | 2240 | 985 | 3 |
| Fin | 2710 | 871 | 202.4 |

The boundary conditions for fluid flow and battery are defined as follows:
- The air velocity is specified at the inlet, while the outlet pressure is set equal to the ambient pressure.
- A no-slip boundary condition is imposed at the interface between the airflow and the battery surface.
- For the thermal analysis, the initial temperature of the battery, air, and environment is set to 298.15 K

In this study, to simplify the simulation, several assumptions are made in the heat transfer analysis like the materials comprising the battery are assumed to be homogeneous, meaning the density, specific heat capacity, and thermal conductivity of each material remain constant throughout the simulation.

The SIMPLE algorithm is employed to solve the pressure-based equations derived from the momentum and mass continuity equations, ensuring coupling between velocity and pressure fields through an iterative solution strategy. A second-order upwind scheme is used to discretize the governing equations.

### 2-4 Phase Change Material Simulation

Fig. 4 shows the PCM geometry of the study, which includes the cell, fins, and housing. The geometry and dimension of PCM are presented in Fig. 5 and Table 4. The energy equation for PCM is described as follows [14]:

$$\frac{\partial(\rho_{pcm}H_{pcm})}{\partial t} + \nabla(\rho_{pcm}uH) = k_{pcm}\nabla^2 T \tag{8}$$

$$H_{pcm} = h + \Delta H \tag{9}$$

$$h = h_{ref} + \int_{T_{ref}}^{T} C_p \Delta T \tag{10}$$

$$\Delta H = \gamma L \tag{11}$$

Where $\rho_{pcm}(kg.,^{-3})$ is the density of PCM, $k_{pcm}$ $(W.m^{-1}.K^{-1})$ is the thermal conductivity of PCM, and $H$ (J) is the enthalpy. $\gamma$ (-) represents the liquid fraction of the PCM, which is calculated iteratively and indicates the volume fraction in liquid form. It ranges from 0 to 1 in the mushy zone. $L$ (J) is the latent heat of the PCM. The material properties of PCM are presented in Table 5.

**Table 4.** Dimensions of PCM system.

| Parameter | Size (mm) |
|---|---|
| $R_3$ | 20 |
| $R_4$ | 19 |
| $H_2$ | 73 |

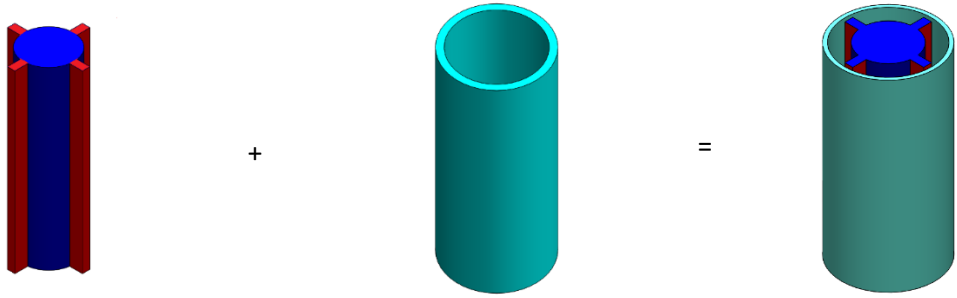

**Fig. 4** Geometry of the BTMS with phase change material and Fins.

During the PCM cooling system the following assumptions are made:

- Laminar flow is assumed due to the very slow movement of the PCM.
- Volume changes during PCM melting are not considered.
- The effect of radiation is neglected.

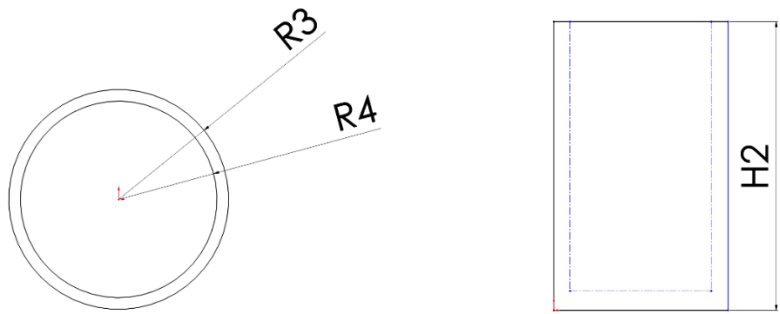

**Fig. 5** 2D schematic of the PCM system with rectangular fins.

Table 5. Properties of phase change material [15].

| Parameter(unit) | Value |
| --- | --- |
| Specific heat capacity $(j.kg^{-1}.K^{-1})$ | 2400 |
| Density $(kg.m^{-3})$ | 769 |
| Thermal conductivity $(W.m^{-1}.K^{-1})$ | 0.146 |
| Viscosity $(kg.m^{-1}.s^{-1})$ | 1.7894e-05 |
| Pure solvent melting heat $(J.kg^{-1})$ | 248000 |
| Solidus temperature (K) | 300.15 |
| Liquidus temperature (K) | 303.15 |

## 2-5 Mesh Generation

Since the mesh independence test is crucial for transient simulations and accuracy of the results, this paper performs a grid independence analysis on the constructed model, as shown in Fig. 6(a-c). The mesh independence study focuses on the average temperature within the battery pack for various grid sizes. It was observed that the temperature varied by only 0.1% when the element size is decreased from 2 mm to 1 mm. Therefore, the element size for this study is selected as 2 mm.

## 2-6 Validation of Numerical Model

To validate the accuracy of the battery simulation, the current numerical model is compared with the work of Reference [13]. The surface temperature of the battery is evaluated against the reference [13] and presented in Fig. 6(d). The results indicate a strong agreement between the numerical simulations and the reference data, confirming the reliability of the model.

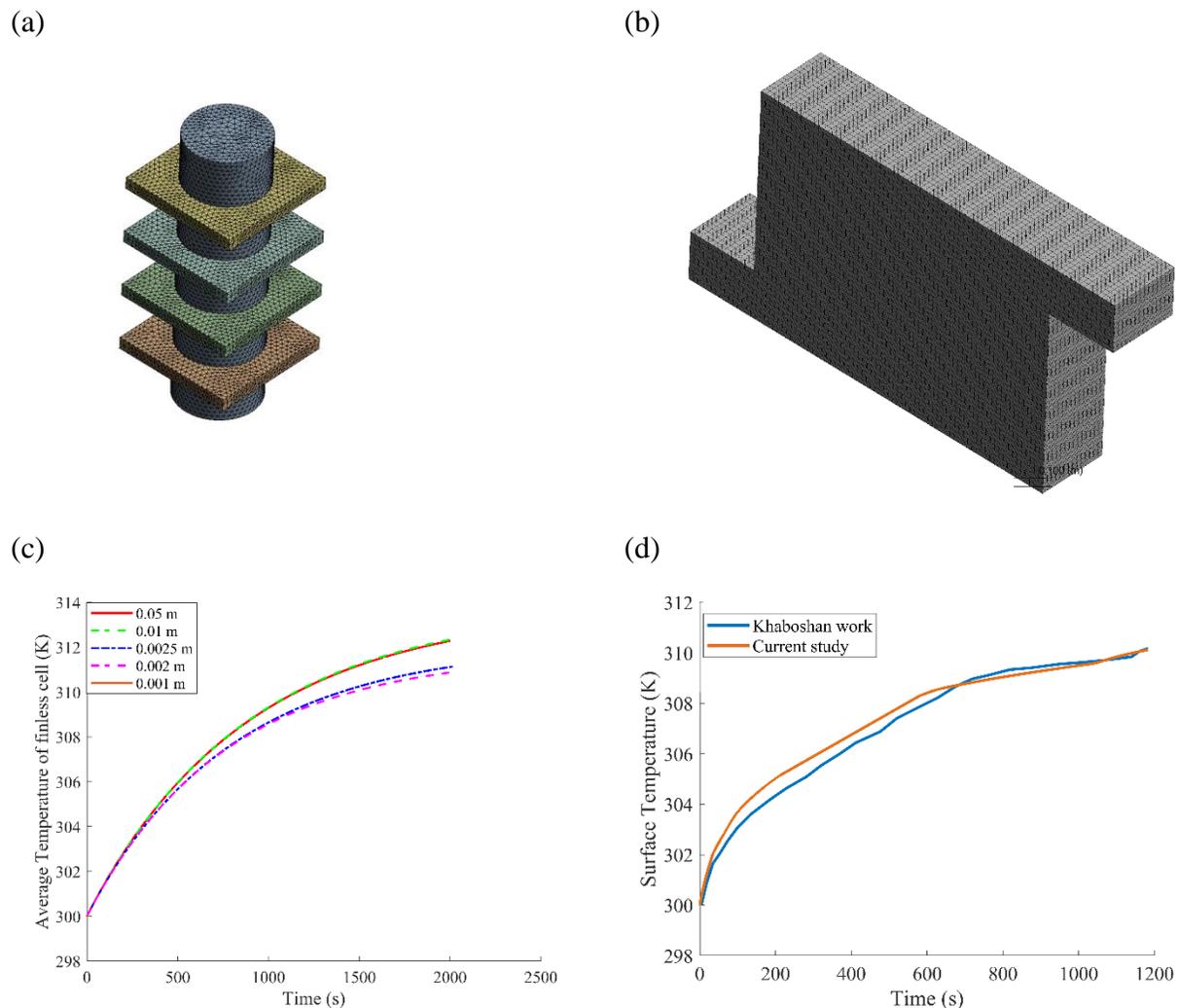

**Fig. 6** Mesh generation for BTMS: (a) Battery with fin, (b) Enclosure, (c) Grid sensitivity analysis with different mesh sizes. (d) Comparison of the present simulation with Ref. [13].

## 3 Results

### 3-1 Air-Cooled-Fin Battery Thermal Management

In this section, the air-cooled BTMS is analyzed which is shown in Fig. 1(a). The airflow velocity is set to 2 m/s, and all results are presented at 2500s. In this simulation, various parameters such as temperature, heat distribution, and the effect of fins on system performance have been analyzed. Fig. 7 compares the battery temperature distribution in an air-cooling system for both the finless battery and batteries with different types of fins. In the finless case, Fig. 7(a), the maximum cell temperature is recorded at 314 K. With circular fins, the maximum temperature decreases to 312 K. When two square fins are used, the maximum temperature drops to 312 K, and with four square fins, the temperature further reduces to 309 K. The lowest maximum temperature of 307 K is observed when rectangular fins are employed. As shown in the contour plot in Fig. 7(b-c), there is no significant difference in the maximum temperature reduction between the two circular fins and the two square fins, as both reduce the maximum temperature by approximately 4.13 K compared to the finless battery. However, incorporating four square and four rectangular fins further reduces the maximum temperature by 6.95 K and 9.5 K, respectively, compared to the finless battery. Thus, the battery with rectangular fins exhibits a 3% reduction in maximum temperature compared to the finless battery in air cooling system.

To maintain the battery temperature difference below 5 K within a single cell, the temperature distribution is analyzed and presented in Fig. 8(a). This figure shows the temperature along the centerline of the battery and across its length. As observed in Fig. 8(a), the battery with fins exhibits a more uniform temperature distribution compared to the battery without fins. The temperature differences along the cell are as follows: 8.8 K for the battery without fins, 4.8 K for circular fins, 4.7 K for two square fins, 4 K for four square fins, and the lowest temperature difference for rectangular fins with 1.03 K. This demonstrates the beneficial effect of fins on the thermal performance of the BTMS. According to this study, longitudinally arranged fins demonstrates superior performance compared to fins arranged squarely. Among different fin shapes, the rectangular fin outperforms circular and square fins due to several factors. The rectangular fin offers a larger surface area for heat transfer, facilitating faster heat dissipation into the surrounding air. Additionally, it promotes a more uniform heat distribution, effectively minimizing the risk of hot spots on the battery surface. Fig. 8(b) illustrates the average integrated temperature of the battery and fins over time. As shown, adding fins reduces the average temperature, with the lowest value observed in the case of four rectangular fins. The results highlight that the use of fins significantly enhances the thermal management system, thereby contributing to improved battery performance and longevity.

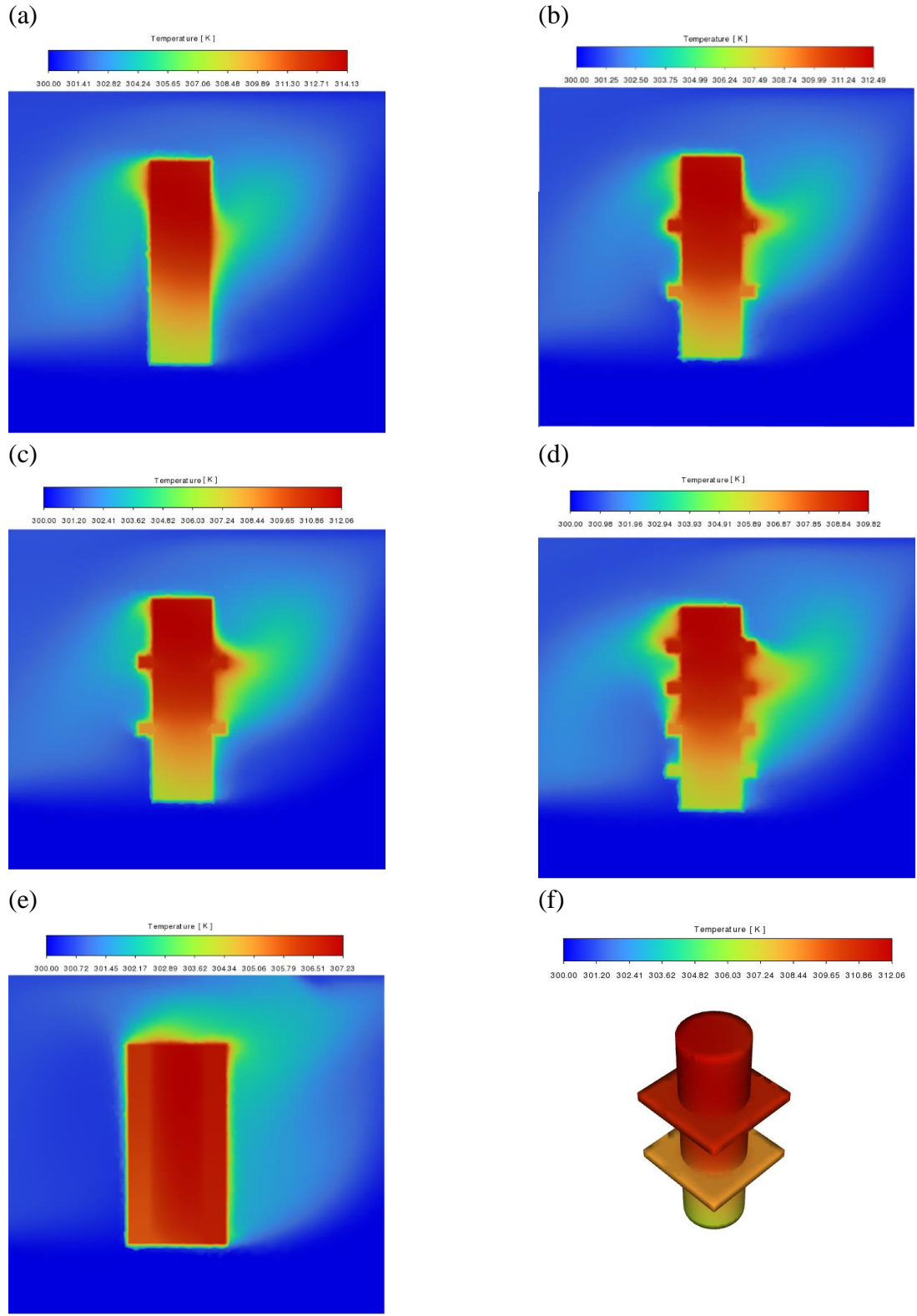

**Fig. 7** Temperature contour of the BTMS in an air-cooled system: (a) finless, (b) with circular fins, (c) with two square fins, (d) with four square fins, (e) with four rectangular fins. (f) Temperature contour of two square fins.

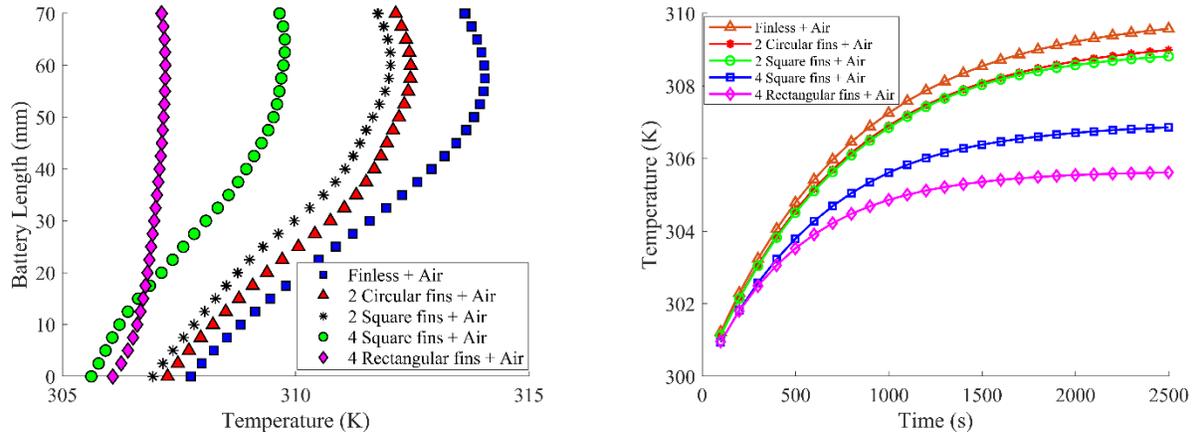

**Fig. 8** BTMS with air-cooled system: (a) Temperature distribution at the center of the battery along its length and (b) Comparison of the average battery temperature over time with different fins.

### 3-2 PCM-Fin Battery Thermal Management

In this section, a single battery cell is studied with a phase change BTMS system, as shown in Fig. 1(b). All results are reported at 2500s. As mentioned, one of the key factors in thermal management is improving heat transfer by increasing the heat exchange area or enhancing the effective thermal conductivity. Based on the previous section, the rectangular fin demonstrates superior performance compared to other fin shapes. Therefore, in this section, rectangular fins are placed around the battery within the phase change material to maximize the heat transfer area and improve thermal regulation.

Fig. 9 shows the temperature contour of the BTMS with PCM for both the finless and four rectangular fin configurations. According to Fig.9, the maximum temperatures are 304 K for the finless system and 302 K for the system with rectangular fins, respectively. It is evident that PCM cooling provides a lower $T_{max}$.

Fig. 10(a) shows the temperature distribution along the length of the battery at the center in the PCM system. The temperature difference along the length is 3.5 K and 1.3 K for the finless and rectangular fins configurations, respectively. The inclusion of PCM significantly enhances the module's temperature uniformity, with only a small increase in temperature. It can be observed that adding fins to the PCM further reduces the temperature difference compared to the finless battery, resulting in improved battery performance. Fig. 10(b) compares the temperature along the centerline of the battery in BTMS using air cooling and PCM. It is observed that the $T_{min}$ is lower in the PCM system compared to the air-cooled system. Additionally, the temperature distribution in the PCM system is more uniform than in the air-cooled system. As the battery temperature increases, the PCM melts and absorbs heat, slowing the rise in $T_{max}$.

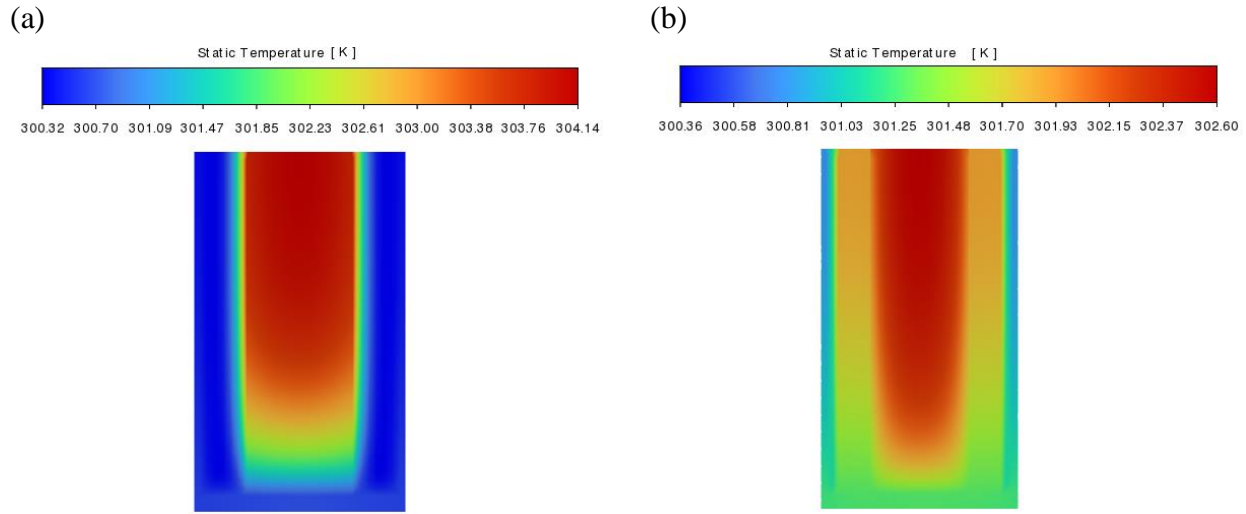

**Fig. 9** Temperature contour of BTMS with phase change material: (a) Finless battery and (b) Battery with rectangular fins.

The phase change occurs more isothermally, resulting in a more gradual temperature increase in the PCM system. The slow rate of temperature increase can be explained by the mechanism of the PCM. Before the PCM starts melting, heat transfer inside it occurs mainly through conduction, while heat storage relies on its sensible heat capacity. As the battery begins operating, its temperature rises rapidly. When the PCM reaches its melting temperature, it begins transitioning from solid to liquid. The melting process happens gradually, starting from the regions closest to the heat source and spreading outward in a radial direction. During the phase transition, heat moves through the PCM via both conduction and convection, while energy storage happens through latent heat absorption. This ability to store heat in the form of phase change helps slow down the battery's temperature rise. The reduction in temperature rise is more noticeable when fins are used, as they increase the contact area between the battery and the PCM, enhancing heat transfer. Using fins in PCM helps transfer heat from the battery surface more effectively by providing additional conductive pathways. This allows the PCM to absorb more heat through its sensible heat storage capacity. However, once the PCM is fully melted, its ability to absorb further heat decreases, and the maximum surface temperature of the battery begins to increase.

So, the fins not only increase the thermal conductivity but also enhance the phase change process by improving the distribution of heat across the PCM. As a result, the temperature rise is more controlled, and the battery remains within a safer operating temperature range for a longer period.

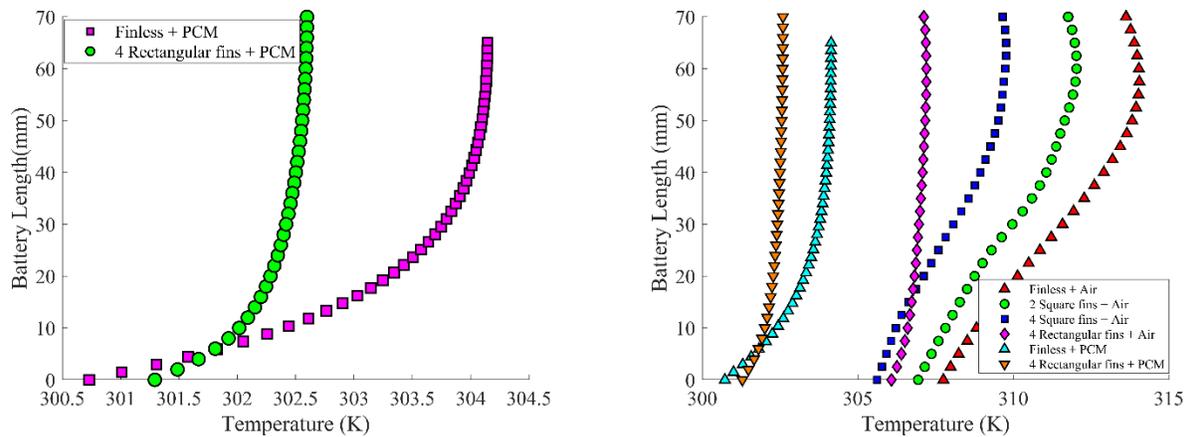

**Fig. 10** Temperature distribution at the center of the battery along its length: (a) Phase change material and (b) Coparing between PCM and air-cooled system.

Fig. 11 presents the liquid fraction in the PCM system. The PCM transitions from solid to liquid around the cell and fins. The presence of fins reduces the volume of PCM. To facilitate a better comparison, the mass of the melted PCM is calculated. The melting rate is lower in the finless battery compared to the finned designs, resulting in a smaller liquid mass of PCM at the end of 2500 seconds for the finless design. The higher melting rate in the finned PCM is attributed to the larger contact area between the finned cell and the PCM, as well as the smaller total amount of PCM in the finned design compared to the finless design.

(a) (b)

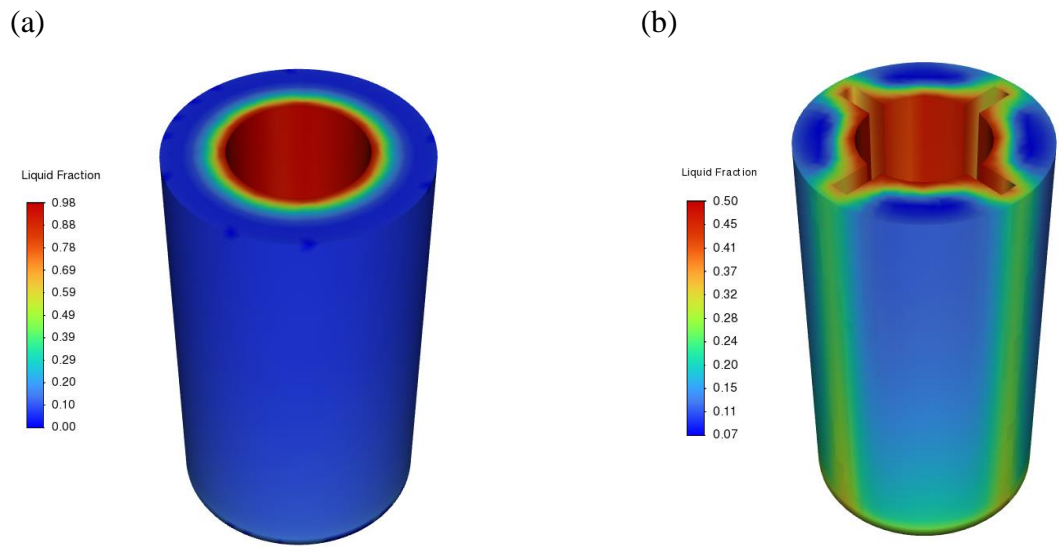

**Fig. 11** Liquid fraction distribution: (a) Finless battery and (b) Battery with four rectangular fins.

## 4 Conclusions

In this study, a comparative analysis is conducted to evaluate the performance of a battery thermal management system using air-cooled and phase change material systems. Several fin designs are proposed and evaluated to enhance BTMS efficiency, and the results of both systems are compared. A three-dimensional numerical simulation is performed using the finite volume method, focusing on key performance metrics such as temperature uniformity, temperature difference, and maximum temperature in both air-cooled and PCM-based systems. The findings of this study are summarized as follows:

- In the air-cooled system, various fin geometries, including circular, square, and rectangular fins, are compared. The results show that by using fins, the maximum temperature decreased from 316 K in the finless battery to 307 K in the rectangular fin design, resulting in a 3% reduction.
- The study demonstrated that longitudinally arranged fins outperformed transversely arranged fins, with rectangular fins providing the most uniform temperature distribution. For example, the temperature difference in the circular fin configuration was 4.8 K, which decreased to less than 1 K with rectangular fins.
- The results revealed that fin structures significantly enhance heat transfer in both systems by increasing the surface area and establishing an effective thermal conduction network.
- In the PCM system, the fins performed similarly to the air-cooled system. By using fins, the maximum temperature was reduced from 304 K to 302 K, improving temperature uniformity. The introduced fin structures create a thermal conduction network within the PCM, facilitating rapid heat transfer from the battery to the PCM and preventing excessive heat accumulation.
- The PCM system also reduced the minimum temperature. The minimum temperature in the air-cooled system was 307 K, which decreased to 302 K in the PCM system.
- The use of PCM resulted in a more uniform temperature distribution along the cell compared to the air-cooled system. The minimum temperature difference in the PCM system is less than 1 K, compared to 1.038 K in the air-cooled system.

## Nomenclature

| | |
|---|---|
| $C_p$ | Specific heat capacity of air ($j.kg^{-1}K^{-1}$) |
| $C_{p,b}$ | Specific heat capacity of battery ($j.kg^{-1}K^{-1}$) |
| $E_{OCV}$ | Open-circuit voltage ($V$) |
| $E_{cell}$ | Cell potential ($V$) |
| g | Gravity ($m.s^{-2}$) |
| H | Enthalpy ($J$) |
| I | Discharge current of the battery ($A$) |
| k | Turbulence kinetic energy ($m^2s^2$) |
| $k_b$ | Thermal conductivity of battery ($W.m^{-1}.K^{-1}$) |
| $k_{eff}$ | Effective thermal conductivity ($W.m^{-1}.K^{-1}$) |
| $K_{pcm}$ | Thermal conductivity of PCM ($W.m^{-1}.K^{-1}$) |
| L | Latent heat of the PCM ($J$) |
| P | Pressure of the continuous phase ($Pa$) |
| T | Air temperature ($K$) |
| $T_b$ | Battery temperature ($k$) |
| t | Time (s) |
| u | Velocity of continuous phase ($m.s^{-1}$) |

**Greek symbols**

| | |
|---|---|
| $\varphi$ | Battery's heat generation rate ($W.m^{-3}$) |
| $\rho$ | Air density ($kg.m^{-3}$) |
| $\rho_b$ | Density of battery ($kg.m^{-3}$) |
| $\rho_{pcm}$ | Density of PCM ($kg.m^{-3}$) |
| $\gamma$ | Liquid fraction (-) |
| $\varepsilon$ | Dissipation rate of turbulent kinetic energy ($m^2s^3$) |
| $\tau_{ij}$ | Reynold's stress tensor ($N.m^{-2}$) |
| $\mu_t$ | Turbulent viscosity ($Pa.s$) |